# Improvement in the Removal Efficiency of the Ultraviolet Laser Ablation by an Additional Simultaneous Irradiation of a Weak Infrared Laser

Y. Kawamura and Akihiro Kai

*Department of Intelligent Mechanical Engineering, Faculty of Engineering, Fukuoka Institute of Technology*
*3-30-1, Wajirohigashi, Higashiku, Fukuoka, 811-0295, Japan*
*E-mail: kawamura@fit.ac.jp*

Remarkable improvement in the removal efficiency of the ultraviolet laser (fourth harmonic wave of Nd:YAG laser) ablation was observed by irradiating a weak infrared laser (fundamental wave of Nd:YAG laser) simultaneously and additionally to various kinds of materials, such as copper, acrylic resin, alumina, silicon and amorphous carbon. The improvement ratio ranged from 50% to 280%, while the laser fluency of the infrared laser was as small as about 1.6% of that of the ultraviolet laser. Although, we have not succeed in explaining the reasons why this effect occurs, it will have the potentiality to decrease the photon cost in the wide range of ultraviolet laser machining, because the method and the phenomena are simple and clear. The improvement in the morphology was also observed, when it is applied to the laser lathe micromachining.

## 1. Introduction

Since 30 years ago, various types of applications using ultraviolet Excimer laser ablations have been studied [1-4]. Among these applications, various kinds of interesting synergistic effect of irradiating two lasers having different wavelength on the material simultaneously have been reported [5-8].

We have been studying laser lathe micromachining using the fourth harmonic wave of Nd:YAG laser (266nm) [9-11], which has almost the same wavelength with that of KrF Excimer laser (248 nm). In these studies, we paid attention to the fact that the laser energy of the fundamental wave of Nd: YAG laser (1.06 $\mu$ m), which was not converted to the higher harmonic wavelength, was not used and simply wasted into the optical dumper. We have been thinking the possibility of the effective utilization of this wasted laser energy, and tried to irradiate a small part of this wasted laser simultaneously with the fourth harmonic wave of Nd:YAG laser (266 nm).

In this report, we describe about the improvement in the ablation efficiency of 50～280％ in the fourth harmonic Nd:YAG laser ablation by irradiating simultaneously and additionally a small amount of the fundamental wave of Nd:YAG laser.

This improvement of the ablation efficiency was observed for various kinds of materials, such as copper, acrylic resin, alumina, silicon and amorphous carbon [12]. It was also observed, even if the laser fluency of the fundamental wave was as small as about 1% of the forth harmonic wave.

The improvement in the morphology was also observed, when this method was applied to "laser lathe machining" [9].

Although, we have not succeed in explaining the reasons why this effect occurs, it will have the potentiality to decrease the photon cost in the wide range of ultraviolet laser machining, because the methods and the phenomena are relatively simple and clear and the forth harmonic wave is always accompanied by the wasted fundamental wave.

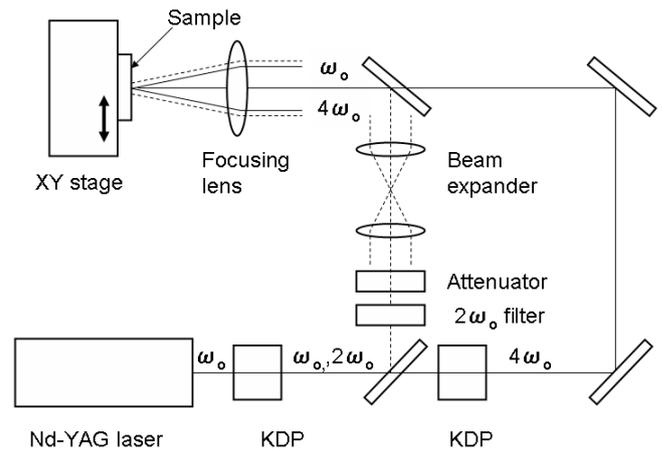

**Fig. 1** Experimental setup for the simultaneous irradiation of UV and IR laser

## 2. Experimental system

Experimental setup is shown in Fig.1. The forth harmonic wave (266nm, UV laser) is generated by a pair of KDP crystals, from the fundamental wave of Nd YAG laser (1.06 $\mu$ m, IR laser). IR laser, which is not converted to the second harmonic wave by the first KDP crystal, is split out from the main beam line by the first beam splitter.

It is expanded by a pair of lenses to adjust the diverging angle, which makes it possible to change the diameter of the focusing spot and irradiation power density on the materials. IR laser is aligned to the same optical axis with UV laser in order to irradiate on the same area of the material surface. The focal length of the focusing lens was 50 mm. The focal spot size of the UV laser was about 15 μm, and that of the IR laser was several times larger, therefore the former is surrounded by the latter.

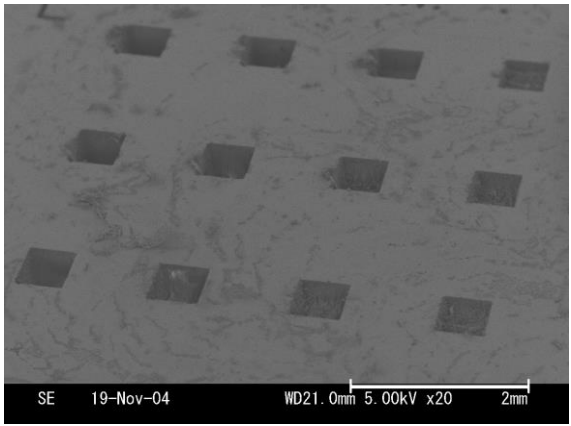

**Fig. 2** Typical example for the laser ablated square regions for the measurement of the ablated depth.

The pulse duration and the repetition ratio of these two lasers (UV laser and IR laser) are about 10 ns and 30 Hz, respectively. The stabilities of the output power of these two lasers are less than 5%. Although, IR laser is irradiated on the material a few nanoseconds earlier than UV laser due to the difference in the optical path length between these two lasers, the irradiation timing can be recognized to be the simultaneous irradiation, because the irradiation pulse duration (10 ns) was longer than the optical delay time.

The focusing points of the laser beam were scanned two-dimensionally on the materials to fabricate a flat top ablated region as shown in Fig.2. The ablation volume for a single laser shot was determined by dividing the removal volume by the total number of shots. The ablation volume was determined by multiplying the depth by the whole machined area. Ablation depth was measured by a surface roughness meter. The ablation depth for a single shot was determined by dividing the ablation volume for a single shot by the scanning unit area.

In order to cover various types of materials, copper, acrylic resin, alumina 96, silicon crystal, and amorphous carbon were chosen.

### 3. Experimental results and discussion

Experimental results of simultaneous irradiation for copper, acrylic resin, alumina, silicon and amorphous carbon are shown using diamonds in Fig. 3, Fig. 4, Fig. 5, Fig. 6 and Fig. 7, respectively. Horizontal scales are the irradiation energy density of IR laser and the vertical scales are the ablation depth for a single shot.

For all of the materials, the irradiation energy density of UV laser was kept to be constant ($3.2 \times 10^{-3}$ J/cm$^2$). The irradiation energy density of IR laser was changed in the range of $5.7 \sim 5.1 \times 10$ J/cm$^2$.

For comparisons, the ablation depth obtained only by IR laser, namely without UV laser, is shown using open circles in these figures.

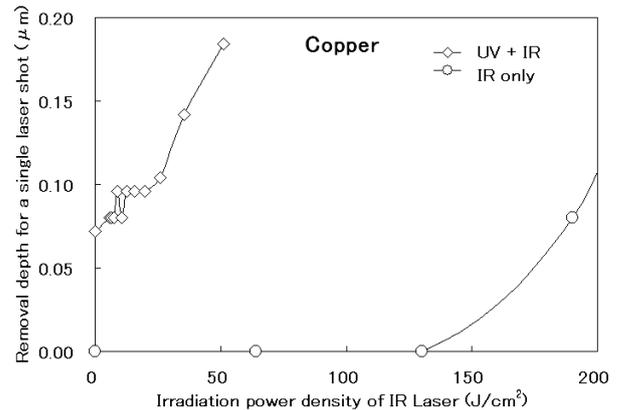

**Fig. 3** Removal depth of copper for a single shot of simultaneous laser irradiation (diamonds) as a function of irradiation power density of IR laser. Open circles show the removal depth by the irradiation of IR only.

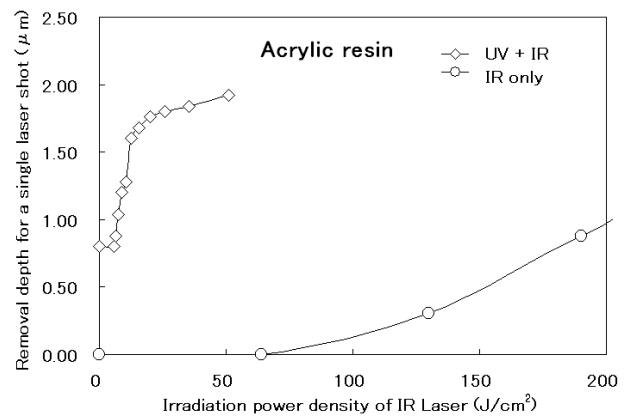

**Fig. 4** Removal depth of acrylic resin for a single shot of simultaneous laser irradiation (diamonds) as a function of irradiation power density of IR laser. Open circles show the removal depth by the irradiation of IR only.

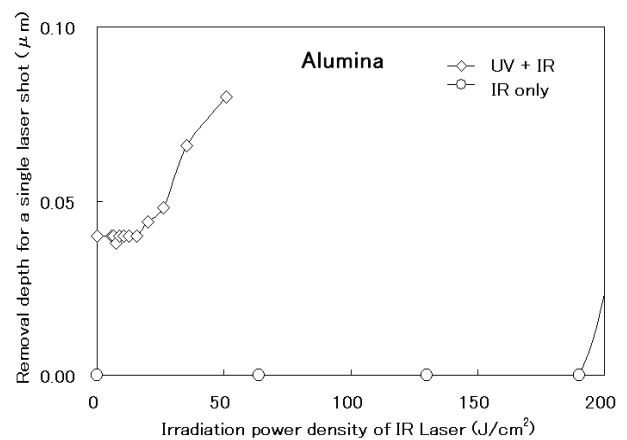

**Fig. 5** Removal depth of alumina for a single shot of simultaneous laser irradiation (diamonds) as a function of irradiation power density of IR laser. Open circles show the removal depth by the irradiation of IR only.

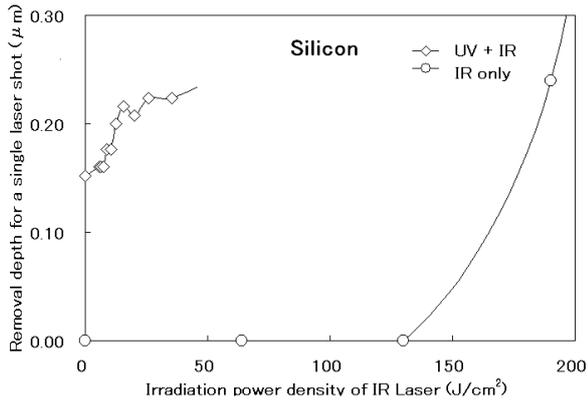

**Fig. 6** Removal depth of silicon for a single shot of simultaneous laser irradiation (diamonds) as a function of irradiation power density of IR laser. Open circles show the removal depth by the irradiation of IR only.

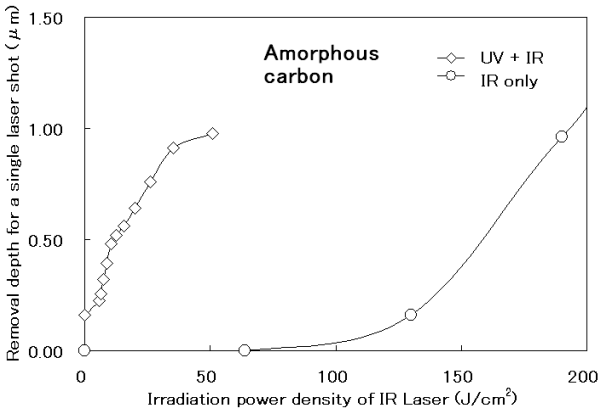

**Fig. 7** Removal depth of amorphous carbon for a single shot of simultaneous laser irradiation (diamonds) as a function of irradiation power density of IR laser. Open circles show the removal depth by the irradiation of IR only.

Although, the irradiation energy densities of IR laser are extremely lower than the ablation threshold of IR laser and, and they are only 0.2% ～1.6% of those of UV laser, the increase in the ablation depth was clearly observed by irradiating IR laser simultaneously and additionally for all the materials.

The increasing ratios were 120%, 130%, 100%, 50% and 280% for copper, acrylic resin, alumina, silicon and amorphous carbon respectively for the irradiation on IR laser power density of 50 J/cm$^2$, which was only 1.6% of UV laser power density ($3.2 \times 10^{3}$ J/cm$^2$).

The saturation of the ablation depth was observed at the irradiation energy density of about 50 J/cm$^2$ for acrylic resin and silicon. It was not observed for copper, alumina and amorphous carbon.

### 4. Discussion

Preheating of the materials caused by the irradiation of IR laser is one of the possible explanations for the phenomenon described here. Even if the irradiation density of the IR laser is less than the threshold density of the IR laser ablation, the surface temperature of the materials increases high enough to change the physical property of the materials, such as breaking stress and optical absorption coefficient for UV laser. The former decreases for the most of the materials, as the temperature increases. The latter is known to increase on the surface of the melted metals. These dual effects is considered to be one of the explanations for the phenomenon.

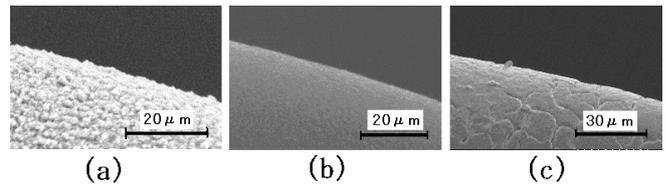

**Fig.8** Typical surface morphology of copper obtained in the micromachining using a laser lathe. (a): rough surface (UV laser only) (b): sooth surface (simultaneous irradiation of UV and IR) (c): scale-like surface (simultaneous irradiation of UV and IR laser)

Fig.8 shows the improvement in the morphology, when the simultaneous irradiation of UV and IR lasers was applied to "laser lath" machining of an axial symmetrical parabolic surface [9].

Fig.8 (a) shows a typical rough surface, obtained for the irradiation of UV laser only. Fig8 (b) shows a typical smooth surface obtained for the simultaneous irradiation of UV laser and IR laser. Fig.9 (c) shows a typical scale-like surface obtained for the simultaneous irradiation of UV and IR lasers, which occasionally appears depending on the irradiation conditions.

In case of the laser lathe machining [9], the finished surfaces are machined by irradiating a focused laser beam parallel to the surface of works, while it is machined irradiating it perpendicular to the surface of the works in case of normal laser machining. In the latter case, this improvement in the morphology was also observed, although it was not so clear as the case in the former.

This improvement in the morphology is considered to be important, when this method is used for the laser micro machining.

### 5. Conclusions

In conclusions, remarkable improvement in the removal efficiency of UV laser (fourth harmonic wave of Nd:YAG laser) ablation was observed by irradiating a weak IR laser (fundamental wave of Nd:YAG laser) additionally and simultaneously for various kinds of materials, such as copper, acrylic resin, alumina, silicon and amorphous carbon. Improvement ratio ranged from 50% to 280%, while the fluency of IR laser was as small as about 1.6% of that of UV laser.

The improvement in the morphology was also observed, when it was applied to the laser lathe machining.

Although, we have not succeeded in explaining the reasons why this effect occurs clearly, preheating of the materials caused by the irradiation of IR laser is one of the possible explanations.

This method will have the potentiality to decrease the photon cost in the wide range of ultraviolet laser machining, because the method and the phenomenon are relatively simple and clear.

The effect described in this paper can be expected to occur for the other various materials, which were not tested in this experiment.